# Studies of $Fe_mIr_n$ (2≤m+n≤4) nano clusters using Density Functional Theory Techniques

S.Assa Aravindh

Theoretical Physics, Institute of Mathematical Sciences, Chennai – 600 113, India

*Email:* assaa@imsc.res.in, mails2asa@gmail.com

**Abstract**

The structure, binding energy, magnetic moments and electronic structure of $Fe_mIr_n$ (2≤m+n≤4) clusters are investigated using state of the art density functional theory techniques. Fully unconstrained structural relaxations are undertaken by considering all possible non equivalent cluster structures. The optimized clusters are all compact, indicating a clear tendency to maximize the number of nearest neighbour Fe-Ir pairs. The binding energy shows an increment with cluster size. All the clusters preserve ferromagnetic order after optimization. The average magnetic moment generally shows an increase with Fe concentration. The spin polarized density of states is largely dominated by the contribution of d orbitals. An important enhancement of the local Fe moments in an Ir rich environment is observed due to the charge transfer between Fe and Ir. On the other hand, the Ir moments are already large in the pure Ir clusters and doesn't show significant enhancement with Fe doping. The HOMO- LUMO gaps show a general reduction with alloying, indicating more metallicity for the doped clusters than the pure ones.

# 1. Introduction

Studies on the structural and magnetic properties of small metal clusters are carried out with great interest, both experimentally [1,2] and theoretically [3,4,5]. These studies are important as clusters play an important bridge between the atom and bulk, thus in understanding the transition from small to big. In small clusters, most of the atoms reside at the surface and hence the structures can be very different from the fragments of bulk phases, and different structures may have quite different magnetic and electronic properties. Owing to the large surface to volume ratio, clusters are more effective catalysts. Apart from the catalytic efficiency, electron localization caused by the small size of the clusters gives rise to larger magnetic moments. Often non-magnetic metals can exhibit magnetism in the cluster form. The high surface to volume ratio of these clusters gives rise to fewer bonds per metal atom and hence frees up non bonded valence electrons, leading to enhanced magnetic moments if left unpaired. Therefore, the magnetic moments in these clusters are very much sensitive to the geometry of the cluster and the chemical environment [6-8]. In this scenario, clusters of transition metals attract particular attention as they are known to display enhanced magnetic moments relative to the bulk counterparts, hence are relevant candidates for magnetic storage devices. They find applications in heterogeneous catalysis too. Among transition metals, Fe clusters receive wide attention due to its distinctive magnetic properties. Small $Fe_N$ clusters are studied both experimentally [9-12] and theoretically [3, 13-18] using different techniques, and showed ferromagnetism with enhanced magnetic moments of about $3\mu_B$/atom and compact structure. . Studies on Fe clusters consisting of 2,3 and 4 atoms carried out by Cheng et al.[3] have shown a ferromagnetic ground state with enhanced magnetic moments and smaller inter-atomic distances compared to the bulk phase. Apart from the clusters of a single element, combinations of transition metals are also studied due to the interesting properties that can arise due to alloying [19]. The binary metal clusters have been shown to exhibit projected magnetic moments that are larger or smaller than the corresponding free metal atoms, depending on the composition of the hetero nucelar cluster [20]. Our interest in this context is devoted to the study of clusters consisting of ferromagnetic 3d and 5d atoms. Presently we focus on clusters of Fe and Ir. Most of the studies on Iridium clusters give emphasis to the catalytic activity and reaction with surfaces [21-22] and the catalytic efficiency of Ir has been used in dissociative chemisorptions. Studies have shown that there exists a low translational energy pathway for the dissociation of methane using Ir as the catalyst [23]. Iridium clusters of various sizes are being studied both experimentally [24, 26] and theoretically [25, 27, 28, 29]. Feng et al [25] investigated the reactivity of Ir clusters from 4 to 10 atoms. Another study on Ir clusters supported on Ir (111) substrate showed that the shape and size affects the stability and higher stabilities are achieved for larger clusters containing 19 and 39 atoms [26]. Bussai et al.'s [27] analysis on the binding energies of 4 atom iridium clusters showed a square planar configuration as the most stable structure using a scalar relativistic variant of density functional theory (DFT) method. Another DFT study of Ir clusters with sizes ranging from n =2 to 64 [28] showed that the magnetic moments and binding energies showed oscillatory behaviour with increase in cluster size. On the other hand, an increase in binding energy with cluster size is also reported in DFT studies [29]. A DFT study involving Fe clusters supported on Ir (111) layers [39] have shown that atomic magnetic moments are induced in the Ir substrate due to the presence of Fe atoms. However, there are very few investigations on how the energetic, stability and magnetic properties vary when Ir clusters are alloyed with another transition metal. In this paper, we present density functional theory calculations of free standing $Fe_mIr_n$, ($2 \leq m+n \leq 4$) clusters. The aim is to investigate how the magnetic moments of clusters will modify when a ferromagnetic metal (Fe) is bonded with a metal that is non-magnetic (Ir) in the bulk phase. Also, Ir is an important catalyst material, and thus investigating the properties of alloyed clusters containing Ir will be interesting. Complete geometry optimization of all the clusters by including all non equivalent combinations and chemical orders will help us to generate the most stable cluster geometries. The magnetic and electronic properties of the optimized geometries are calculated. The computational methodology is presented in the next section and the results and discussions are presented in section 3. The results are compared with other theoretical and experimental data wherever possible. Finally our conclusions are presented in section 4.

## 2. Methodology

First principles density functional theory calculations are carried out for Fe-Ir clusters using the plane wave density functional theory program PWSCF (ESPRESSO Version 5) [30]. The Rappe-Rabe-Kaxiras-Joannopoulos (RRKJ) ultra soft pseudo potential with non-linear core correction [31] is used for Fe and Ir with the Perdew-Burke-Ernzerhof (PBE) generalized gradient approximation (GGA) correction [32] formalism employed for the exchange-correlation functional. Both of these pseudo potentials are obtained from the PWSCF pseudo potential reference [33] and the atomic electronic configuration of Fe and Ir atoms used in the pseudo potential are $3d^7 4s^1$ and $5d^7 6s^2$ respectively. The main difficulty in the theoretical study of binary-metal clusters is the identification of the low energy configurations accurately as there exist diverse geometrical conformations and distributions of the different atoms in the cluster. Various ordered and disordered structures may exist with different degrees of intermixing and segregation. Hence, all possible initial geometries and topologies with all possible distributions of the Fe and Ir atoms should be taken into account. These structures are then need to be relaxed in fully unconstrained manner until the equilibrium configuration is reached. In the present study concerning $Fe_m Ir_n$ ( $2 \leq m+n \leq 4$), we have considered all possible topologies and chemical order of the Fe and Ir atoms within the cluster including the pure $Fe_m$ and $Ir_n$ limits. We have generated all possible geometries as described in Ref [20]. Even though different initial geometries are considered, after the unconstrained relaxations the energetically stable ones are considered for further analysis and presented in the paper. The computational parameters such as the size of the super cell, the kinetic energy cut off, the cut off for the electron density and the force convergence threshold are checked for convergence. The clusters were placed in a simple cubic super cell of size 20x20x20 Å$^3$, assuring that the interaction between the clusters and their replicas in neighbouring cells are negligible. This also enables us to consider only the Gamma (Γ) point ($k$=0) when integrating over the Brillouin zone. The real space grid for the numerical calculations involving the electron density is described by an energy cut off of 250 Ry. All the clusters are fully relaxed until the inter atomic forces are smaller than Z 10$^{-3}$eV/Å. The stability of each cluster is assessed based on the binding energy per atom. The binding energy per atom is calculated as,

$E_B$ = [$mE$(Fe) + $n$ $E$(Ir) – $E$(Fe$_m$ Ir$_n$ )] / N,

where $E$(Fe) and $E$(Ir) are the total energy correspond to the $m$ Fe and $n$ Ir isolated atoms. The binding energy of the cluster (in eV/atom) is defined as the difference in total energy between the interacting atom system and the free atom system. Hence clusters with more binding energy are more stable and vice versa. To aid the convergence, Marzari Vanderbilt smearing method [34] is used, with a Gaussian broadening of 0.001 Ry. The average as well as the local magnetic moments is calculated by spin polarized total energy calculations. Localized electron population analysis is carried out by integrating the projected density of states, to calculate the local charges and local magnetic moments on the atoms. The electronic structure of the clusters is calculated to understand the effects of alloying on the cluster properties. It is to be noted that in this study, only the atomic spin magnetic moments are computed, i.e. spin orbit interactions are not taken into account. An initio total energy calculations in small Fe clusters [7] has shown that relaxation of the clusters leads to stabilization of collinear magnetic structures. In another study by Longo et al, also the non collinear magnetic order is not found to be existing in Fe-Mn clusters [8]. In view of the above results a non- collinear magnetic structure is not expected for the Fe-Ir clusters studied in the present work and the calculations are done in the collinear frame work. The energy difference between the highest occupied molecular orbital (HOMO) and the lowest unoccupied molecular orbital (LUMO) are calculated and this energy gap is a used to understand the stability of the clusters. The HOMO-LUMO energy gap $E_g$ (in units of eV) is a characteristic quantity of the electronic structure of the cluster and is a measure of the stability of the clusters to undergo activated chemical reactions with small molecules. The magnitude of HOMO-LUMO gap may vary with both the size and composition of the cluster and the manner in which these orbitals are getting filled determines the properties of the clusters.

## 3. Results and Discussion

In this section, the investigations on the ground state structure, stability and magnetic properties of $Fe_mIr_n$ clusters are described. The main focus is to understand the effects of size and chemical composition on the stability and magnetism. The similarity and differences in properties between clusters of various sizes are enumerated. The geometry of the clusters after optimization and their properties are summarized in tables 1, 2 and 3. Before proceeding with the cluster calculations, we have calculated the lattice parameter and cohesive energy of bulk Fe and Ir to test the accuracy of computational scheme. The calculated (measured) lattice constant and cohesive energy of bulk and Fe and Ir are 2.87 Å (2.86 Å), 4.41 (4.28 eV), 3.85 Å (3.84 Å), and 6.90 eV (6.94 eV) [35] respectively. The good agreement with experimental values show that the computational method is appropriate to investigate the Fe-Ir clusters.

We have started our calculations with the Fe dimer, and then replaced one Fe atom by Ir to understand the change in properties with alloying. It is seen that the binding energy increases from Fe-Fe to Ir-Ir, indicating that the bonding in $Fe_2$ is the weakest and the bonding resulting from the Fe-Ir pair is stronger than the Fe-Fe bonds. The binding energy obtained for $Fe_2$, 1.05eV (using GGA in the present study) turns out to be larger than the experimental value, (0.65 eV/atom) [11] but is in agreement with that reported in other theoretical studies [4, 13, 14]. This over estimation of binding energy compared to experimental results may be related to the choice of exchange and correlation functional employed. The bond length analysis also indicates the same trend as that of binding energy. The Fe-Fe bond length, 2.00 Å, is in consistent with the experimental values 1.87Å [9] and 2.02Å [10] obtained in different studies and theoretical results of Gong et al, which has shown a bond length of 2.09 Å [4]. However, the calculated bond length is considerably smaller than the nearest neighbour distance, 2.48 Å in bulk bcc Fe [35]. On close analysis it can be seen that the bond length follows the trend of the atomic radii (atomic radii of Fe being 1.72 Å, whereas that of Ir is 1.87 Å [29], such that $d_{Ir-Ir} > d_{Fe-Ir} > d_{Fe-Fe}$. The trend of the $Fe_2$ and Fe-Ir to be more compact compared to the $Ir_2$ indicate a more metallic character of the 3d bonding and a tendency of a more covalent and directional bonding in the case of 5d ones. The HOMO- LUMO energy difference of the dimers are large, indicating a lower reactivity, however the alloying seems to reduce the gap. The average magnetic moment varies from 3.38 $\mu_B$ to 2.51 $\mu_B$ as one goes from $Fe_2$ to $Ir_2$, indicating nearly full polarization of all the d electrons. The magnetic moment per atom for $Fe_2$, 3.38 $\mu_B$ is in agreement with other DFT results [3, 14,15], and is in keeping with the value obtained in experiments (3.3±0.5) [12]. For the pure Fe and Ir dimers, the local and average magnetic moments are equal and it indicates that the spin density, $m(\vec{r}) = n_\uparrow(\vec{r}) - n_\downarrow(\vec{r})$, is quite localized around the atoms. The difference between the average and local magnetic moments give a measure of the spin density $m(\vec{r})$ which is spilled off to the immediate environment of the atoms. The local magnetic moment of Fe and Ir in the mixed Fe-Ir dimer is quite remarkable, particularly if one compares them with the pure dimers. The Fe local moment in Fe-Ir, is 3.53 $\mu_B$ is 0.15 $\mu_B$ larger than Fe$_2$ and even larger than the Fe atom, while the Ir moment is reduced by a similar amount (0.14 $\mu_B$) as can be seen from table 1. It can also be seen from the table that the electronic charge around Fe decreases where as that of Ir increases. The reduction in Ir magnetic moment is attributed to the transfer of d electrons from Fe to Ir, which enhances the number of d holes and allows the Fe atom to develop a significantly larger spin moment. This increase in Fe magnetic moment occurs at the expense of Ir atom and this observation is qualitatively in agreement with the larger Pauling electro negativity of the Ir atom ( $\chi_{Fe}$ = 1.83, $\chi_{Ir}$ = 2.20) [37].

Even though we have started our calculations by considering both linear and triangular geometry for the trimers, after optimization, the most stable geometry turned out to be an equilateral triangle and this is true for all compositions. The linear geometries relaxed into triangular geometries, favouring a higher coordination number and the results of the stable configurations for each composition are presented in table II. One observes that the inter atomic distances follow the trends in the atomic radii as in the case of dimers. For Fe$_3$, the lowest energy structure is an equilateral triangle with bond length 2.19 Å, in good agreement with previous DFT studies, which have reported bond lengths of 2.15Å, 2.37Å, 2.04Å, 2.10Å, 2.14Å and 2.11Å [References 3,4, 14, 13, 15, and 16 respectively]. Compared to Fe$_2$, the bond length shows a tendency to increase, indicating that the system always favours the highest dimension for longer bond length. The Fe rich trimers tend to be more compact indicating more metallic character of bonding as in the case of dimers. The binding energy shows a monotonous dependence on composition similar to the case of dimers, indicating that the Ir-Ir bonds are the strongest in the trimers too. However, the trend of increase in binding energy with increase in number of Ir atoms indicates that stability increases with increase in number of Fe-Ir bonds. It is to be noted that $FeIr_2$ is more stable than $Fe_2Ir$ because Ir-Ir bonds are in general stronger than those between Fe atoms. The HOMO –LUMO band gap shows a monotonous decrease with Ir composition, indicating more chemical activity. The average magnetic moment of $Fe_3$, 3.45 $\mu_B$/atom, is in agreement with previous DFT studies [4, 5, 13-15] and close to the experimental results [12]. The ground state of $Ir_3$ is equilateral with local as well as average magnetic moment of 2.22 $\mu_B$/atom. This indicates that the spin polarization is dominated by electrons occupying localized states and the spill off contributions is not important in the case of $Ir_3$, unlike the other trimers. It is to be noted that the local magnetic moments in the trimers always show a ferromagnetic coupling. For mixed compositions, the presence of Fe-Ir bonds enhances the local Fe magnetic moments beyond 3 $\mu_B$/atom. The substitution of one Ir atom in $Fe_3$ to yield $Fe_2Ir$ increases the local Fe magnetic moment to 3.56 $\mu_B$/atom even though the average magnetic moment shows a monotonous decrease with Ir addition. Similar to the dimers, this is mainly due to the charge transfer from Fe to Ir. It can be seen that, quantitatively, the local Fe and Ir magnetic moments are similar to the values calculated in dimers. In addition to this, local magnetic moments of Ir atoms in the mixed dimers are enhanced compared to pure $Ir_3$ cluster. This indicates the significance of the proximity of Fe on the magnetic properties of Ir atoms.

The tetramers are also studied in detail as they are the smallest clusters with a three dimensional geometric structure. We have inspected different possible isomers of the Fe, Ir and Fe-Ir tetramers, such as tetrahedral, rhombus, square planar and a linear chain structure, each of them having different coordination number. In the tetrahedral structure, the coordination number of all atoms is 3, while in the rhombus, two atoms have coordination number 3 and two atoms have 2. In the square planar structure, all atoms have coordination number 2. Despite considering all non equivalent cluster geometries for each composition, after complete geometry optimization, the most stable ones turned out to be tetrahedrons. In this structure, the tetrahedral topology ensures the freedom of permuting the Fe and Ir atoms within the cluster without altering the chemical order. The most stable isomer for the $Fe_4$ cluster is a regular tetrahedron with average magnetic moment of $3.55\mu_B$, and is comparable to previous studies [4], which showed the stable structure to be a regular [13,16] or distorted tetrahedron [14, 15]. The obtained bond length of $Fe_4$, 2.34 Å is in keeping with other DFT results [3, 4]. The increasing trend of bond length from $Fe_2$ to $Fe_3$ to $Fe_4$ indicates that the clusters always prefer the highest dimension for longer bond length and maximum pairs of nearest neighbour bonds. Previous studies on of $Ir_4$ clusters show contradictions regarding the ground state structure. The predicted stable geometries in theoretical studies include tetrahedral [25], a butterfly structure [27], rhombus [28] and square [29] whereas experiments has showed tetrahedral as the most stable structure having a bond length of 2.71Å [38]. We have seen that the tetrahedral structure is the most stable structure in agreement with this experimental result and theoretical study which obtained tetrahedron having bond length of 2.79 Å [25]. For the alloyed tetramers also, the tetrahedron structure came out to be the most stable of all the configurations studied. The binding energy exhibits a non-monotonous dependence on composition, unlike the situation in dimers and trimers. This shows that with the increase in number of Ir atoms, the Fe-Ir bond strength increases. Considering the change in binding energy from $Fe_4$ to $Ir_4$, the relative stability of the clusters can be correlated to the number of homogeneous and heterogeneous bonds. It can be understood by looking at the most stable composition, $FeIr_3$, which has three Fe-Ir and Ir-Ir bonds. Coming to $Fe_2Ir_2$, replacing one Ir by one Fe implies replacing two Ir-Ir bonds by a relatively stronger Fe-Ir bond and a weaker Fe-Fe bond. In this way, the binding energy is not altered significantly. A visible decrease in binding energy is obtained only for larger Fe concentration. We have noticed that the average bond length increases from 2.0 Å for the Fe dimer to 2.34 Å for the Fe tetramer, indicating a tendency to approach the nearest neighbour distance in the bulk bcc Fe with the increase in cluster size. The calculated bond length (2.44 Å) and the magnetic moment/atom (2.13 $\mu_B$) of Ir$_4$, is comparable to that obtained in Ref.[29] (2.31 Å and 2.00 $\mu_B$ respectively), however they have obtained octahedral geometry as the stable structure. Yet another study showed the square planar structure with large binding energy [28] while we have observed that the square planar structure upon complete geometry optimization transforms in to a tetrahedral geometry. These variations in stability can be attributed to the difference in approximations involved in the choice of pseudo potential and exchange correlation functional. The HOMO – LUMO energy gap shows a reduction in with alloying favouring more metallicity in the alloyed clusters. Turning to the magnetic properties, a ferromagnetic order is observed in all the clusters and the average magnetic moment linearly increases with Fe concentration. This increase in magnetic moment can be related to the increase in bond length with cluster size. We have noticed that with the increase in compactness, magnetic moment decreases. This is due to the fact that, in compact clusters, the number of nearest neighbours is larger than more open structures, and to have large magnetism, more energy is required. Hence the compact structures tend to possess small magnetic moments. For the tetramers also, the presence of Fe in the neighbourhood enhances the Ir local moments considerably. The average as well as local magnetic moment of pure Fe tetramer, 3.55 $\mu_B$/atom is larger than obtained for the Ir$_4$ clusters, contrary to the zero magnetic moment reported for the tetrahedral structure by T. Pawluk et al [28].

It is very interesting to investigate, at least for some of the clusters, how the electronic structure changes with respect to alloying. To this aim, the spin polarized electron d- density of states of the Fe-Ir tetramers as well as that of the pure Fe$_4$ and Ir$_4$ clusters are calculated and shown in the figure 1. The total density of states of the clusters is also shown along with for comparison. The s and p orbital contributions are found to be negligible. It is seen that the majority and minority density of states are not equal, indicating a ferromagnetic ordering in the clusters. The electronic distribution of the d orbital shows localized nature for both pure Fe and Ir tetramers and it is mainly distributed around the Fermi level. For the pure $Ir_4$ clusters a dominant d electron contribution occur near $E_F$ with the characteristic exchange splitting between the spin up and spin down states. It can be seen that the total and d- orbital density of states are hardly distinguishable in $Ir_4$ cluster. The valence spectrum mainly consists of these d-electron contributions. Four peaks are visible for the $Ir_4$ cluster 5d orbital, all of them corresponding to the valence band. The majority spin bands are almost completely filled and the Fermi level lies at the prominent peak in the minority spin channel. The effective d band width in the pure Ir tetramer is about 1eV larger than the pure Fe tetramer. On the other hand, for the $Fe_4$ cluster the majority spin bands are completely filled with the highest majority state lying about 0.5eV below the Fermi level ($E_F$). It can be inferred from these observations that $Fe_4$ is a strong ferromagnet with saturated magnetic moments while $Ir_4$ can be considered as a weak unsaturated ferromagnet. The density of states of the alloyed clusters reflects the cross over between the behaviour of $Fe_4$ and $Ir_4$ tetramers. For $FeIr_3$, where one Ir atom is substituted by Fe, we can still find states near to the Fermi energy. Even though the Fermi energy shows a tendency to approach the top of the majority band, the magnetic moments are not saturated. It can be noticed that the majority d bands of Ir is close to the Fermi energy and the Fe doping doesn't seem to reduce the d-bandwidth significantly. However, the peak structure changes as a consequence of Fe doping. Coming to the equal concentration, $Fe_2Ir_2$, the exchange splitting increases. In the majority band, Ir dominates Fe at the higher energy region near to Fermi level, while Fe d states dominate the bottom of the band. In the minority band, the contribution of Ir decreases below $E_F$ and is larger above $E_F$ contrast to the Fe behaviour. Consequently, the local Ir moments are reduced compared to the Fe moments. In the Fe-rich $Fe_3Ir_1$ cluster the band width of majority spin states are narrowed and become comparable to that of pure Fe tetramers. For the pure $Fe_4$ cluster, the exchange splitting increases and the majority spin band is almost saturated. It is observed that the large local magnetic moments of Fe originates from the majority d bands located between -5 to -1 e V below the $E_F$.

## 4. Conclusions

In summary, investigations on the structure, binding energy, magnetic moments and electronic structure of $Fe_mIr_n$ (2≤m+n≤4) clusters were performed using density functional theory techniques. All the clusters are fully relaxed to their ground structures. The optimized geometries are in general compact, with a strong tendency to intermix to maximize the number of heterogeneous Fe-Ir bonds. Even though the bond length increases with the increase in size for the pure Fe and Ir clusters, it is far from the bulk nearest neighbour distance of Fe (2.48 Å) and Ir (2.71 Å) [35]. The magnetic order in the optimized clusters is found to be ferromagnetic. The average magnetic moment showed an overall increment with increase in cluster size for the pure Fe clusters, but remained in the vicinity of 2.2 $\mu_B$ for the pure Ir clusters. The trends in magnetism are controlled to a large extent by the Fe content in the alloyed clusters. The spin polarized density of states of the tetramers is largely dominated by the contribution of d orbitals. An important enhancement of the local Fe moments in an Ir rich environment is observed. On the other hand, the Ir moments are already large in the pure Ir clusters and hence doesn't show significant enhancement with Fe doping. The HOMO- LUMO gap shows a general reduction with alloying, indicating more metallicity for the doped clusters than the pure ones.

**Acknowledgement**

The author thank Institute of Mathematical Sciences (IMSc) , Chennai, India for the computational and financial support.

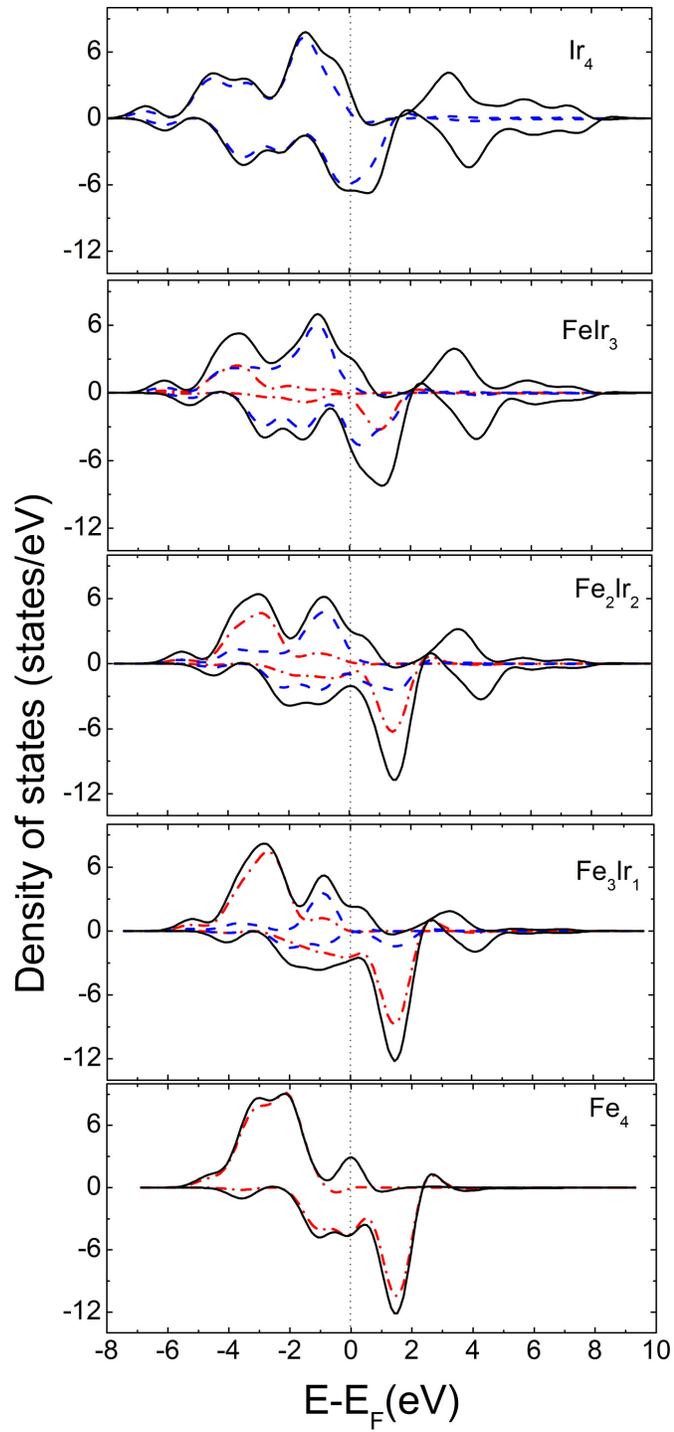

Figure1. The spin polarized electronic density of states (DOS) of Fe-Ir tetramers. The total (solid) DOS as well as the Fe projected (dash-dot) and the Ir projected (dash) d-electron DOS are shown. The majority and minority spin are represented by positive and negative values respectively. The vertical line at zero corresponds to the Fermi level.

Table I. The structure, calculated average inter atomic distance (d) , binding energy ($E_B$), the HOMO-LUMO energy gap ($E_g$), the electronic charge (q) around Fe or Ir atoms ordered from top to bottom, the average spin magnetic moment (μ), and the local spin magnetic moment at the Fe or Ir atoms, ($μ_{Fe}$ /$μ_{Ir}$ ) are shown for the dimers.

| Cluster | $d_{(m-n)}$ (Å) | $E_B$ (eV) | $E_G$ (eV) | charge | $\mu$ ($\mu_B$) | $\mu_{Fe}$ ($\mu_B$) | $\mu_{Ir}$ ($\mu_B$) |
|---|---|---|---|---|---|---|---|
| Fe$_2$ 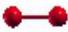 | 2.00 | 1.05 | 0.89 | 7.94 | 3.38 | 3.38 | |
| Fe-Ir 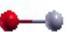 | 2.13 | 1.76 | 0.69 | 7.51 9.37 | 2.89 | 3.53 | 2.37 |
| Ir$_2$ 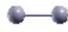 | 2.26 | 1.97 | 0.90 | 8.98 | 2.51 | | 2.51 |

Table II. The structure, calculated average interatomic distance between atoms m(Fe) and n (Fe or Ir), binding energy ($E_B$), the HOMO-LUMO energy gap ($E_g$), the electronic charge (q) around Fe or Ir atoms ordered from top to bottom, the average spin magnetic moment (µ), and the local spin magnetic moment at the Fe or Ir atoms, ($µ_{Fe}$/$µ_{Ir}$) are shown for the trimers.

| Cluster | $d_{(m-n)}$ (Å) | $E_B$ (eV) | $E_g$ (eV) | charge | µ ($µ_B$) | $µ_{Fe}$ ($µ_B$) | $µ_{Ir}$ ($µ_B$) |
|---|---|---|---|---|---|---|---|
| Fe$_3$ 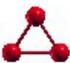 | 2.19 | 1.29 | 0.91 | 7.93 | 3.45 | 3.47 | |
| Fe$_2$Ir 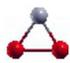 | 2.28  2.25 | 1.98 | 0.78 | 7.49  9.75 | 3.16 | 3.56 | 2.25 |
| FeIr$_2$ 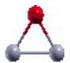 | 2.26  2.43 | 2.38 | 0.32 | 7.26  9.29 | 2.67 | 3.47 | 2.25 |
| Ir$_3$ 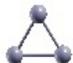 | 2.42 | 2.45 | 0.03 | 8.98 | 2.22 | | 2.22 |

Table III. The structure, calculated average inter atomic distance between atoms m(Fe) and n (Fe or Ir) ordered from top to bottom as $d_{Fe-Fe}$, $d_{Fe-Ir}$, $d_{Ir-Ir}$, binding energy ($E_B$), the HOMO-LUMO energy gap ($E_g$), the electronic charge (q) around Fe or Ir atoms ordered from top to bottom, the average spin magnetic moment (µ), and the local spin magnetic moment at the Fe or Ir atoms, ($\mu_{Fe}/\mu_{Ir}$) are shown for the tetramers.

| Cluster | $d_{(m-n)}$ (Å) | $E_B$ (eV) | $E_g$ (eV) | charge | µ (µ$_B$) | $\mu_{Fe}$(µ$_B$) | $\mu_{Ir}$ (µ$_B$) |
|---|---|---|---|---|---|---|---|
| Fe$_4$ 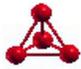 | 2.34 | 1.87 | 0.89 | 7.91 | 3.55 | 3.55 | |
| Fe$_3$Ir 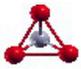 | 2.39<br>2.36 | 2.42 | 0.55 | 7.51<br>10.09 | 3.27 | 3.57 | 2.21 |
| Fe$_2$Ir$_2$ 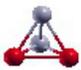 | 2.41<br>2.38<br>2.56 | 2.73 | 0.28 | 7.28<br>9.59 | 2.91 | 3.55 | 2.24 |
| FeIr$_3$ 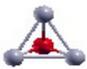 | 2.40<br>2.55 | 2.86 | 0.32 | 9.24<br>7.11 | 2.63 | 3.56 | 2.31 |
| Ir$_4$ 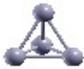 | 2.44 | 2.74 | 0.91 | 8.98 | 2.13 | | 2.13 |